# Study of monolithically integrated 940 nm AlGaAs distributed Bragg reflectors on graded GaAsP/bulk Si substrates

Jia Guo,[1] Yunlong Zhao,[1,*] Markus Feifel,[2] Hao-Tien Cheng,[3] Yun-Cheng Yang,[4] Lukas Chrostowski,[5] David Lackner,[2] Chao-hsin Wu,[3,4] and Guangrui (Maggie) Xia[1,**]

*Jia Guo and Yunlong Zhao contributed equally to this work.

**Corresponding author

**Abstract**: We report the fabrication of monolithically integrated 940 nm AlGaAs distributed Bragg reflectors (DBRs) on graded GaAsP/Si substrates. Low-density surface bumps and cross-hatch patterns were observed on the DBR surfaces. Cross-sectional DBR layers are smooth and flat. The reflectance spectra of the GaAsP/Si DBRs have lower intensities than the GaAs DBRs and have double peaks. Transfer matrix method calculations, surface scratch and polishing tests were conducted, which suggest that the surface cross-hatch was the cause of the inferior DBR reflectance spectra.

## 1. INTRODUCTION

Vertical-cavity surface-emitting lasers (VCSELs), consisting of two distributed Bragg reflectors (DBRs) with an active region in between for light emission perpendicular to the wafer surfaces, are widely used optical components in optical networks, parallel optical interconnects, laser printers, high-density optical disks and more and more importantly in sensing [1]. The market size of VCSELs is forecasted to increase from 1.2B USD in 2021 to 2.4B USD in 2026 at a compound annual growth rate (CAGR) of 13.6% owing to the fast-growing demand from the mobile/consumer product manufacturers, the data communication companies, and the automotive industries [2][3]. Most VCSELs are fabricated with III-V semiconductors like GaAs and AlGaAs with emission wavelengths from 650 to 1300 nm. To address the dramatically increased demand for VCSELs, making them monolithically on much larger Si substrates, typically 8" and 12" in diameter, has potential advantages of much higher throughput, and lower cost compared with those on conventional 4" and 6" GaAs substrates [4]–[8].

To date, heterogeneous integration approaches such as wafer bonding techniques have been examined and demonstrated to be effective to integrate VCSELs with Si substrates. In those methods, VCSELs are still prepared on GaAs substrates and then transferred to Si substrates by either direct wafer bonding techniques [9], [10] or adhesive bonding techniques [11]–[13].

In contrast to heterogeneous integration, monolithic integration, also known as direct growth, of VCSELs on bulk-Ge or Si substrates is of great interest because of the simpler process and larger substrate size, thus potentially much larger throughput, and lower cost. Very promising results have been obtained on the bulk Ge substrates recently. Bulk-Ge-based 940 nm VCSELs by monolithic integration were first demonstrated and shown to have comparable performance and smaller wafer bow compared with the conventional bulk-GaAs counterparts in 2021 [7]. In our recent work published in 2022 [8], GaInAs/bulk-Ge-based 940 nm DBR performance was on par with that of the bulk-GaAs DBRs. Compared with bulk Ge substrates, typically 8" in diameter, Si substrates can be even larger, typically 12" in diameter, enabling higher throughput and has the capability to integration with other Si photonic devices. It is well-known that the material properties mismatch between Si and GaAs/AlGaAs, including lattice size mismatch, crystal structure difference, and coefficient of thermal expansion (CTE) mismatch, can generate high-density defects and residual stress [14], which are the roadblocks for the monolithic integration of GaAs-based devices on Si.

Growing GaAs monolithically on Si has been studied for several decades with only limited success such as the case in Si-based multijunction solar cells [15], [16], and aspect ratio trapping (ART) Ge on Si-based DBRs [17]. One natural question is whether Si-based VCSELs are still meaningful for industry applications due to the high defect density. Our considerations are the following: 1) common VCSELs are few-micron-diameter lasers with oxide layer confinement for apertures, which have more immunity to threading dislocations (TDs). TDs can be leakage current paths and diffusion paths. The oxide layers formed after the VCSEL epitaxy may effectively interrupt TDs originally through the oxide areas. Therefore, only TDs through the apertures matter; 2) many VCSEL array modules need to be paired with diffusers to illuminate a wide field of view. This means that non-working VCSELs only decrease the overall power output without sacrificing functionality. Therefore, the study of Si-based VCSELs is still meaningful. The focus of this paper is on GaAsP/bulk Si-based 940 nm n-DBRs, which is the first 940 nm Si-based DBRs studied. The change of focus from the previous ART-Ge/Si substrates in [17] was mainly due to the discontinued fabrication by the original grower. The choice of 940 nm DBRs was due to the higher popularity of 940 nm VCSELs in 3D sensing.

A common method to overcome the materials properties mismatch is to introduce buffer layers and use offcut substrates. For example, Ge, with the similar lattice constant to GaAs and AlAs (less than 0.1% lattice constant difference), was examined as the buffer layer between Si substrates and AlGaAs DBRs [17]. Lin et al. reported a successful integration of AlGaAs DBRs on an on-axis Si (100) substrate with a polished ART-Ge buffer layer with offcut characteristics [17]. High aspect ratio trenches were used to trap most of the threading dislocations in Ge and the offcut characteristics of the Ge reduce antiphase domains.

Another approach is to introduce graded buffer layers, such as $Si_xGe_{(1-x)}$ [18] and $GaAs_xP_{(1-x)}$, to achieve high-quality GaAs layers on Si substrates for the subsequent GaAs-based devices. The Si substrates with graded $GaAs_xP_{(1-x)}$ have been designed and used for multijunction solar cells on offcut (100) Si substrates reported by Feifel et al. [15], [16]. High-quality GaP nucleation layers with similar lattice constant to Si (around 0.4% lattice constant difference) have been adapted as excellent templates for the epitaxial growth of GaAsP [19]. Additionally, 14 layers of graded $GaAs_xP_{(1-x)}$ provide a smooth transition from the GaP nucleation layer at the bottom to the high-quality GaAs layer at the top [16]. Beyond the lattice constant mismatch, CTE mismatch leads to crack formation during cooling. Based on our previous study of 850 nm AlGaAs VCSEL epitaxy on GaAs/ART-Ge/Si substrates conducted in 2018 [20], the average crack space was around 100 μm.

In this work, GaAs/graded GaAsP/offcut (100) Si substrates were investigated and compared with bulk-GaAs and GaInAs/bulk-Ge substrates. GaAs/graded GaAsP/offcut (100) Si wafers were chosen due to the lower defect density (TDD around $2.2 \times 10^7$ $cm^{-2}$) in the top GaAs layers and the proven success of using them as the substrates for multi-junction solar cells producing high AM 1.5 solar cell efficiency of 22.3% [16]. Our first step towards 940 nm Si-based VCSELs was to build 940 nm bulk Si-based AlGaAs DBRs. The DBR reflectance spectra were measured. Material analysis, 1-dimensional (1D) transfer matrix method (TMM) simulations, as well as surface treatment tests, were conducted to understand the impacting factors and the improvement methods of the DBR stopband characteristics.

## 2. DBR GROWTH EXPERIMENT, RESULTS AND DISCUSSION

### 2.1 Epitaxial growth

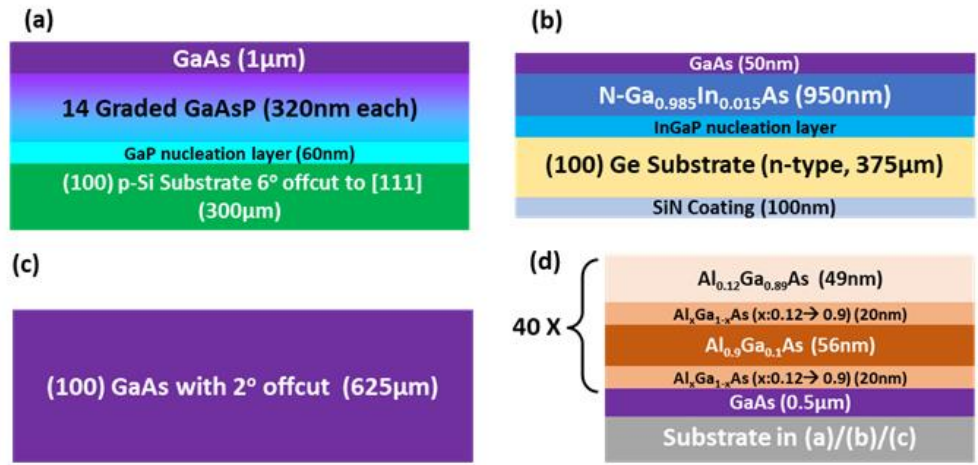

*Figure 1: Schematic illustration of: (a) a GaAs/graded-GaAsP/Si substrate, (b) a GaAs/GaInAs/bulk-Ge substrate, (c) a control GaAs wafer and (d) DBR fabricated on a substrate shown in (a)/(b)/(c). The 0.5-micron GaAs is the base layer before the DBR growth.*

The schematic structures of the GaAs/graded GaAsP/Si substrate, GaAs/GaInAs/bulk-Ge substrate, GaAs wafer, and the DBR grown on a bulk-GaAs substrate are shown in Fig. 1. For simplicity, in the following discussions, these substrates are referred as GaAsP/Si, GaInAs/bulk-Ge and bulk-GaAs substrates with the corresponding top epitaxy GaAs layers and the nucleation layers omitted unless otherwise noted.

For the GaAsP/Si substrates in Fig. 1 (a), a 300 μm thick p-type (100) Si wafer with a 6° offcut towards [111] was used. The metallorganic chemical vapor deposition (MOCVD) was performed in a CRIUS close-coupled shower-head reactor from AIXTRON. The Si substrates, before loaded into the MOCVD chamber, were dipped in diluted HF to remove the native oxide. The 60 nm thick GaP nucleation layers were then deposited based on the two-step nucleation strategies as described in Ref. [19][21]. After that, the sample was withdrawn and loaded again into another chamber for graded GaAsP buffer layers deposition at 700 °C. Triethylgallium (TEGa), arsine ($AsH_3$), phosphine ($PH_3$), and silane ($SiH_4$) were used as precursors for 14 layers of GaAsP with the thickness of 320 nm for each layer, resulting in a grading rate of 0.8 %/μm. Finally, a 1 μm GaAs layer was deposited as the virtual substrate for the subsequent DBR growth. The detailed fabrication processes of GaInAs/bulk-Ge substrate in Fig. 1 (b) was described in Ref. [14].

Before the DBR growth at LandMark Optoelectronics Corporation in an MOCVD reactor with 3" wafer pockets, the 4" GaAsP/Si substrates were cut into pieces of 1 cm x 1 cm. A few 3" 375 μm thick Si dummy wafers with 1 cm x 1 cm square holes made by laser machining were used as adapters to fix the wafer pieces in place in the 3" wafer pockets. After cleaning with diluted HF acid, a 0.5 μm GaAs base layer was grown to create a fresh and clean surface for the subsequent 940 nm DBR growth.

The DBR shown in Fig. 1. (c) consists of 40 periods of 20 nm $Al_xGa_{1-x}As$ (x: 0.12 → 0.9) / 56 nm $Al_{0.9}Ga_{0.1}As$ / 20 nm $Al_xGa_{1-x}As$ (x: 0.9 → 0.12) / 49 nm $Al_{0.12}Ga_{0.88}As$ superlattice with 3 x $10^{18}$ $cm^{-3}$ n-type doping, which was designed to have center of peak reflectivity at 940 nm wavelength. The DBR growth temperature was at 760 °C and the pressure was 50 mbar. $H_2$, $SiH_4$, $AsH_3$, and $Ga(CH_3)_3$ (TMG) gases were used as the precursor for MOCVD of GaAs. Additional $Al_2(CH_3)_6$ (TMA) was introduced as the source of Al during the growth of $Al_xGa_{1-x}As$. The same DBR growth process was applied on a 3" 625 μm thick 2-degree offcut (100) GaAs wafer as the control sample.

## 2.2 DBR surface characterizations

The surface properties of the DBRs grown are essential for the optical performance of DBRs and VCSELs. In this work, optical microscope, scanning electron microscope (SEM) as well as atomic force microscope (AFM) analysis were conducted for the DBRs on the four different substrates.

Fig. 2 shows the optical microscope images of DBR surfaces with 10 times magnification. Stress-induced cross-hatch pattern is seen on GaAsP/Si DBR in Fig. 2 (a), and no surface bumps are observed. The GaInAs/bulk-Ge DBR in Fig. 2 (b) has a very clean surface similar to the bulk-GaAs DBR in Fig. 2 (c).

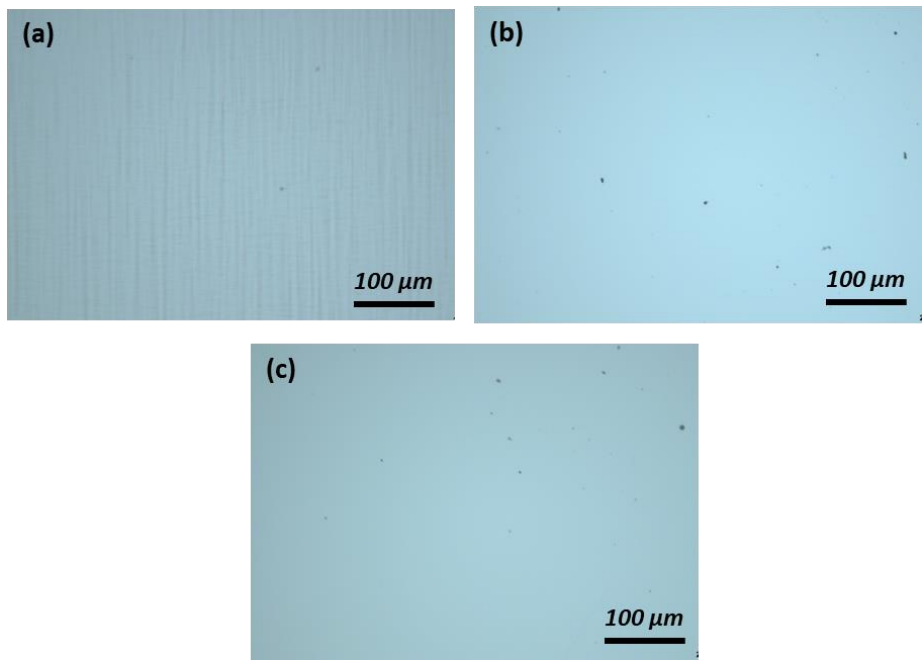

*Figure 2: Optical microscope images of (a) GaAsP/Si DBR, (b) GaInAs/bulk-Ge DBR and (c) bulk-GaAs DBR. The magnification is 10 times.*

To further analyze and quantitatively compare the surface condition of DBRs, SEM images were collected using an FEI Quanta 650 SEM with a back-scattered electron (BSE) detector. The operation voltage was set to be 20 kV and the working distance was around 10 mm. As seen in Fig. 3, surface bumps (indicated by the red arrows) and cracks (indicated by the yellow arrow) are observed on GaAsP/Si DBR. The surface bump density and crack spacing averaged over a total imaging area of 0.0107 mm$^2$ are 39.2 mm$^{-2}$ and 138.4 µm respectively for GaAsP/Si DBR. GaInAs/bulk-Ge DBR surface is free of cracks and only one bump was seen within a large scanning area which translates to the surface bump density of 5.3 mm$^{-2}$. Control Ge DBR is free of surface bumps and cracks. The GaAsP/Si DBR and GaInAs/bulk-Ge DBR have surface bumps but the density is small. The estimated defected surface area percentages for these two DBRs are 1.57 and 0.21 %, which will translate to yield losses. If the surface bumps only impact the yield, with larger substrates, this loss is small and negligible. However, as shown in 2.3, these surface bumps and cross-hatch degrade the reflectance.

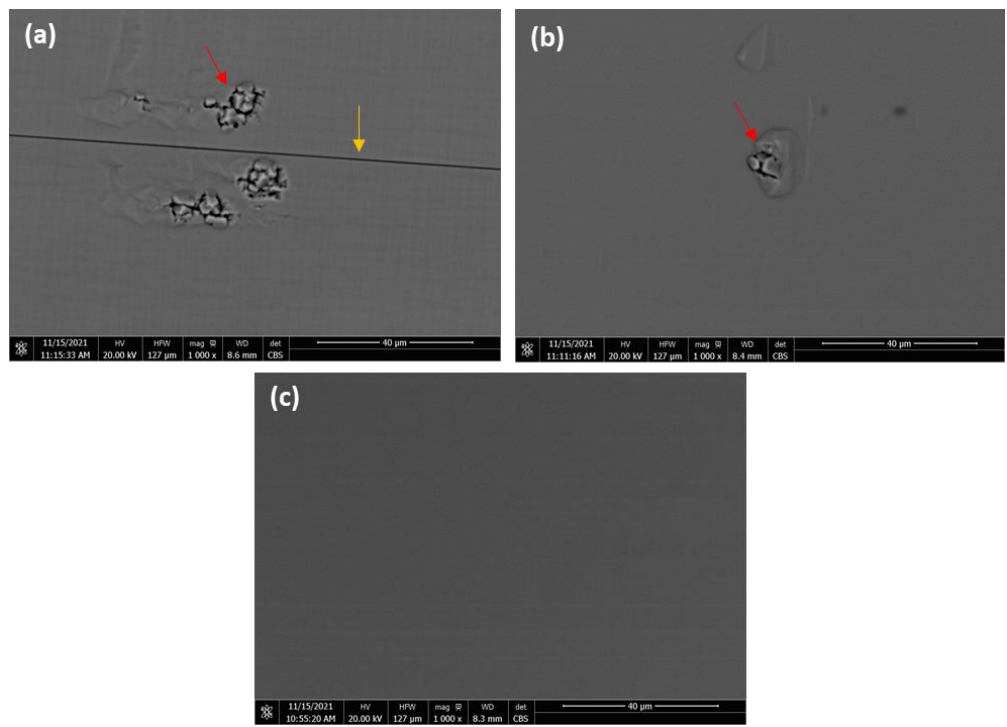

*Figure 3: Surface SEM images of (a) GaAsP/Si DBR, (b) GaInAs/bulk-Ge DBR and (c) bulk-GaAs DBR. The magnification is 1000 times.*

AFM measurement was then conducted on a Bruker's Dimension Icon AFM to investigate the surface roughness of the DBRs. A 50 μm X 50 μm region was scanned with a scan rate of 0.2 Hz and a line density of 512 imaged point /scanned lines. The root mean square (RMS) roughness measured for GaAsP/Si, GaInAs/bulk-Ge, and bulk-GaAs DBRs are 8.52, 0.77, and 0.26 nm respectively. The GaAsP/Si DBR has highest RMS surface roughness, consistent with the optical and SEM observations. A micron-scale cross-hatch pattern was observed on the GaAsP/Si DBR, and a nano-scale cross-hatch pattern was observed on GaInAs/bulk-Ge DBR.

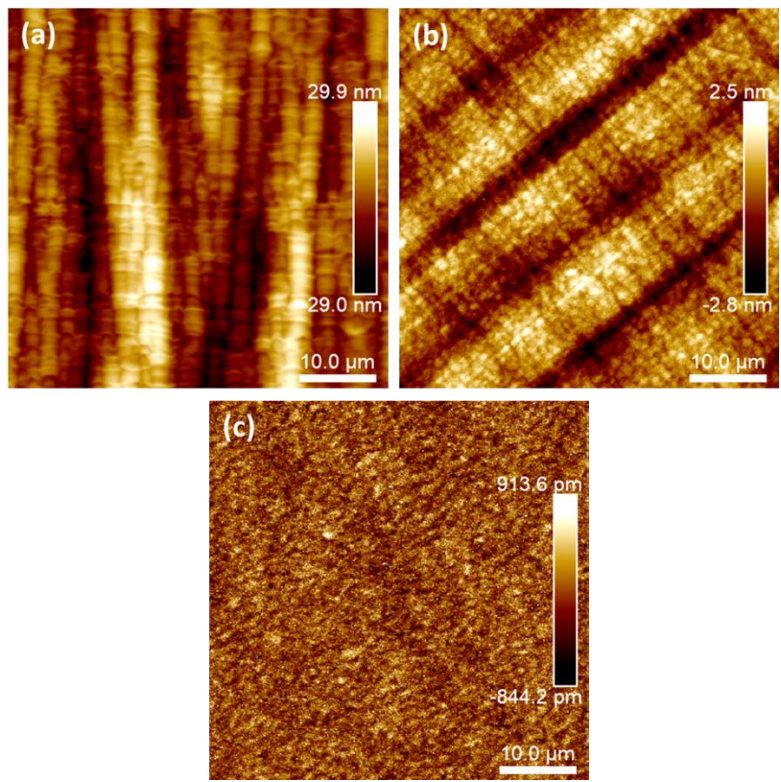

*Figure 4: AFM images of (a) the GaAsP/Si DBR, (b) the GaInAs/bulk-Ge DBR and (c) the GaAs DBR.*

The surface characterization results are summarized in Table 1 for easier comparison.

*Table 1: Summary of the surface characterization results.*

| Substrates | Optical | SEM | RMS roughness by AFM |
|---|---|---|---|
| GaAsP/ 300-μm-Si | Cross-hatch pattern with cracks | Bump density: 39.2 mm$^{-2}$ Crack spacing: 208.8 μm | 8.52 nm, μm-scale cross-hatch |
| GaInAs/ 375-μm-bulk-Ge | Clean without cracks | Bump density: 5.3 mm$^{-2}$ | 0.77 nm, nm-scale cross-hatch |
| 500 μm-bulk-GaAs | Clean | Clean | 0.26 nm |

## 2.3 Reflectance spectrum

The normal-incidence reflectance spectra of DBRs were collected with a Filmmetrix F20 thin film analyzer at room temperature. The measurements were calibrated to include the substrate thickness difference and substrate surface tilt effect. The results were normalized to the maximum reflectance of bulk-GaAs DBR, which is shown in Fig. 5. There is a noticeable blue shift on the reflectance spectrum of all samples when compared with the control GaAs DBR. The stopband center of the bulk-GaAs DBR is located at 942.94 nm wavelength, while the centers of stopband for GaAsP/Si DBR and Ge DBR are at 905.55, and 930.16 nm. This is a result of thinner DBRs compared with the bulk-GaAs DBR revealed by the cross-section SEM shown in Fig. 5. As mentioned above, the DBR growth process is optimized for 3" GaAs wafers at Landmark. Unlike the bulk-GaAs DBR grown on the whole 3" GaAs wafer, the GaAsP/Si DBR, and GaInAs/bulk-Ge DBR were grown on a 1 cm x 1 cm substrate fixed in dummy Si adapters. This may have caused a different and uneven heat conduction and precursor gas flow, which lead to the slower and non-uniform growth rates. Nevertheless, the stopband widths are comparable with the values of 76.06, 74.37, and 77.95 nm for GaAsP/Si DBR, bulk-Ge DBR, and bulk-GaAs DBR respectively. The maximum reflectance of the bulk-Ge DBR is 100.1 % of that of the bulk-GaAs DBR. However, the two peaks of the GaAsP/Si DBR spectrum are only at 72.24 and 88.42% of the maximum reflectance of the GaAs DBR, and the stopband dip is at 66.41%. This may be explained by the higher surface defect density and surface roughness [22]. Other than the lower reflectance at the stopband, the GaAsP/Si DBR shows an unexpected reflectance spectrum with a center dip and two peaks instead of a flat stopband. Our first speculation was that the DBR layer thickness might be off-target and/or not consistent, which was checked with the cross-section SEM discussed below.

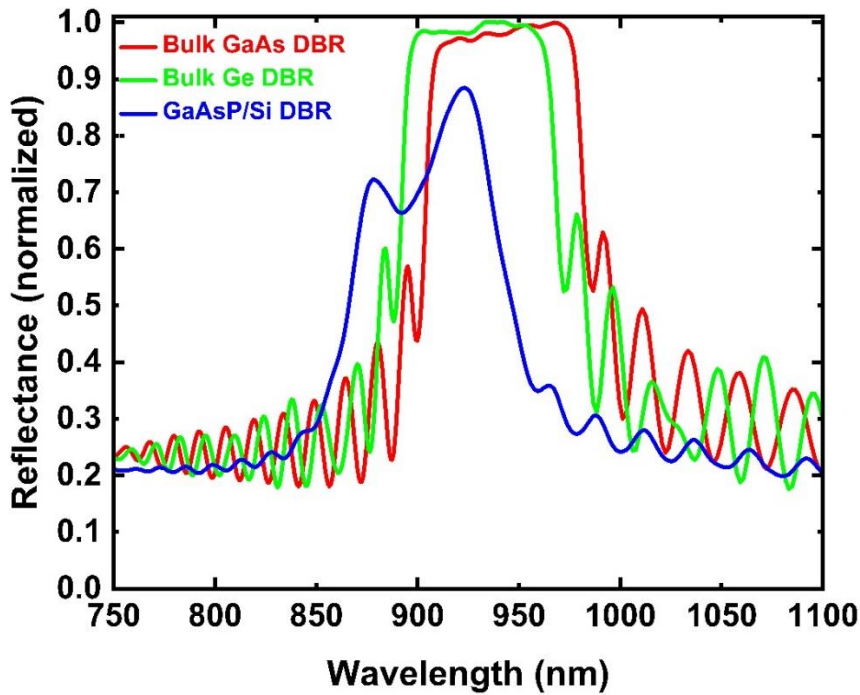

*Figure 5: Normal-incidence reflectance spectra of DBRs grown at each substrate. All spectra are taken at the center of each sample normalized by the maximum GaAs DBR reflectance value.*

**2.4 Cross-section SEM**

Cross-section SEM imaging was conducted to measure the thickness of each DBR period. SEM samples were prepared by manually cleaving the grown DBR with a diamond scribe to obtain a fresh cross-section. FEI Nova NanoSEM system was used with the immersion mode and gaseous analytical detector (GAD). The operation voltages used were 10 kV and the working distance was 5–6 mm. Fig. 6 shows the example cross-section SEM images of GaAsP/Si DBR captured at the bottom (upper left), middle (upper right), and top (lower left) of the DBR. To minimize the inaccuracy brought during thickness measurement and the value reading, two periods of DBR $n_1/n_2$ layers were measured each time.

Fig. 6 (d) summarises the double-period cross-section SEM thickness results, which clearly shows the bulk-GaAs DBR layers are thicker compared with the other three type substrates. Bulk-Ge DBR and bulk-GaAs DBR thickness are quite uniform and comparable with the standard derivations of 2.5 and 3.3 nm respectively. The GaAsP/300-μm-Si DBR has the largest thickness standard derivation of 9.5 nm. The bottom DBR layers are thinner than the top DBR layers with the maximum thickness difference of 40 nm. The high DBR thickness variation of GaAsP/Si DBR can be explained by the sample placement and the shadowing effect. During the DBR growth, 1 cm x 1 cm GaAsP/Si substrate was inserted into a 3" Si adaptor with some narrow air gap around it. The thickness of GaAsP/Si is 305.5 μm in comparison to the 375 μm thickness of the Si adaptor, which may have led to non-uniform precursor gas flow and results in thinner DBR with high variation. While the GaAs/GaInP/375-μm-bulk-Ge/SiN is about 376 μm thick, very close to the Si adapter thickness of 380 μm. The heat conduction of Si substrates is also expected to be different from Ge and GaAs substrates, which may result in a slight growth temperature dependence. Table 2 summarizes the double-period DBR layer thickness by cross-section SEM and the reflectance results.

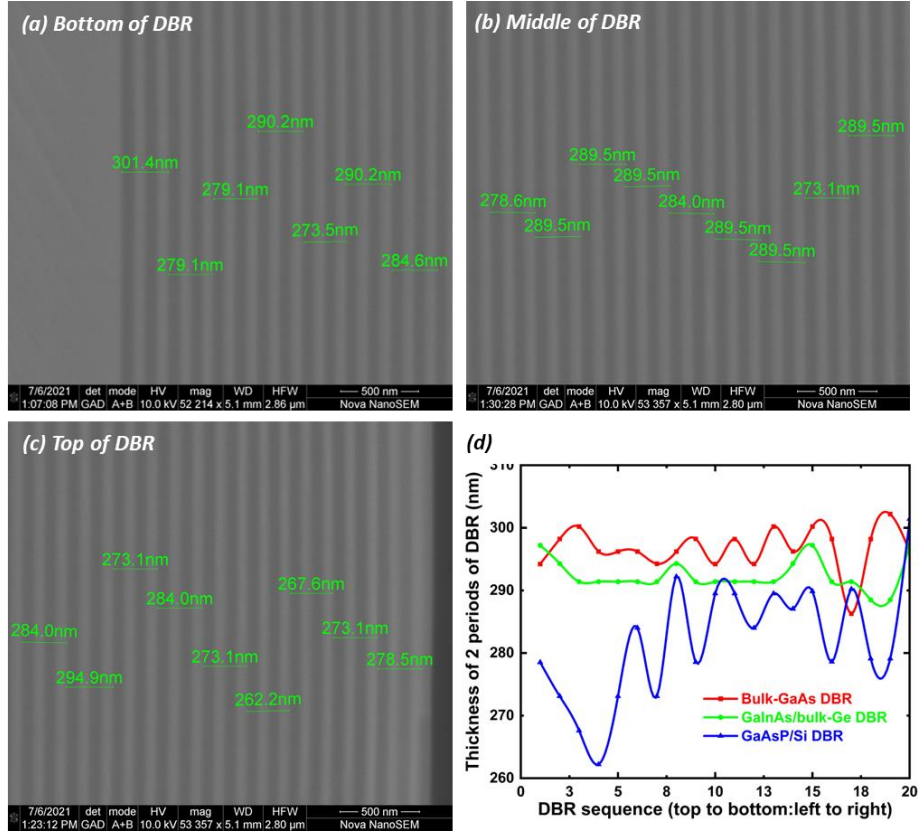

Figure 6: Cross-section SEM image of GaAsP/Si DBR measured at bottom (top left), middle (top right) and top (bottom left) of DBR. Comparison of collected double-period DBR layer thickness on different substrates (bottom right).

Table 2: Summary of the DBR layer thickness by cross-section SEM and reflectance results.

| Substrate | Average 2-period DBR layer thickness (nm) | Standard deviation (nm) | Stopband characteristics | | |
|---|---|---|---|---|---|
| | | | Shape | Width (nm) | Center (nm) |
| GaAsP/ 300-µm-Si | 282.0 | 9.5 | two peaks with a dip | 76.06 | 905.55 |
| GaInAs/ | 292.4 | 2.5, thinner than the GaAs | Flat | 74.37 | 930.16 |

| 375-μm-bulk-Ge | | counterpart by 1.5% | | | |
| 500 μm-bulk-GaAs | 296.7 | 3.3 | Flat | 77.95 | 942.94 |

## 3. TMM DBR SIMULATIONS

Are the stopband differences a result of the one-dimensional (1D) epitaxy layer thickness or the 3D surface roughness? With the thickness data obtained from the SEM analysis as the inputs, we first did 1D simulations of the reflectance spectrum of DBRs using the transfer matrix method (TMM), as shown in Fig. 7. The refractive indexes of $Al_{0.9}Ga_{0.1}As$ and $Al_{0.12}Ga_{0.88}As$ initially used were 3.0342 and 3.4795. These values were then adjusted slightly to 3.038 (+0.1%) and 3.389 (-2.6%) respectively in order to match the experiment result of control GaAs DBR. This slight adjustment to the refractive indexes is acceptable due to process variation. A reasonable matching between the simulation and the measured spectra was obtained on Ge DBR with the adjusted refractive indexes as well. The simulation result of the GaAsP/Si DBR is significantly different from the GaAs DBR with a narrower flat stopband and two strong satellite peaks beside the stopband. However, there is still a flat stopband around 940 nm wavelength, which does not match the measured double-peak-shaped reflectance spectrum. The limitation of TMM simulation is that only the DBR thickness along the z-axis (direction perpendicular to DBR surface) was considered assuming that there is no thickness variation in the in-plane directions, i.e. the x and y directions. This was an oversimplified assumption.

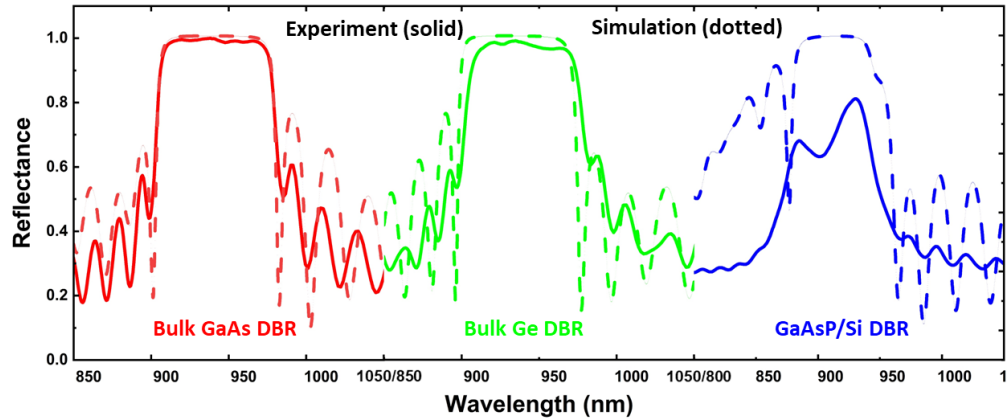

*Figure 7: Simulated reflectance spectrum of 3 DBRs (dotted line) and experimental spectrum (solid line).*

## 4. SURFACE SCRATCH TEST

As the unexpected double-peak-shaped stopband of the GaAsP/Si DBR could not be explained by either the surface defects or the z-axis thickness uniformity, we suspected that the unique and periodic epitaxy-induced cross-hatch pattern might be the root cause. However, there was no study to reveal the impact of the cross-hatch pattern on DBR performance to the best of our knowledge. Thus, we came up with a simple surface scratch test idea to introduce a cross-hatch pattern to the originally perfectly clean and flat GaAs DBR surface to see if the double-peak-shaped stopband could be reproduced from the scratched GaAs DBR.

### 4.1 Formation of the cross-hatch pattern by epitaxy

The cross-hatch surface morphology is commonly observed in the heteroepitaxial thin-film system consisting of two materials with different lattice constants. In our case of GaAs epitaxial film on (100) Si wafer, the crosshatch pattern is in the direction of [011] and [0$\bar{1}$1] [23]. One of the proposed formation mechanisms of the cross-hatch pattern is a combination of misfit dislocation generation and glide process [24]–[26]. Initially, the plastic relaxation process, which takes place in the strain film (GaAs in our case), leads to the misfit dislocation, and causes the formation of surface steps. Subsequent mass transport that eliminates the surface steps results in the crosshatch pattern.

**4.2 Experiment of surface scratch test**

The basic experiment setup is demonstrated in Fig. 8. A commercial silicon carbide (SiC) sandpaper with a grit number of 10000 (equivalent to an average particle size of 2.11 µm) was used to create scratches on the originally perfectly clean and flat GaAs DBR surface. Firstly, the GaAs DBR sample was placed on a rigid surface. During the die saw process of this sample from a whole GaAs wafer, the four edges were cut along the crystal direction [110] and [$\bar{1}$10]. Then the sandpaper was placed on top of the GaAs DBR with the abrasive surface contacting with the sample. Five pieces of microscope glass slides were covered on top of sandpaper to ensure close contact between the sample and sandpaper and to provide a constant pressing force. Finally, sandpaper was gently pulled out to create scratches on the sample surface along crystal direction [110] and [$\bar{1}$10]. The same process was repeated multiple times with fresh sandpaper used every time.

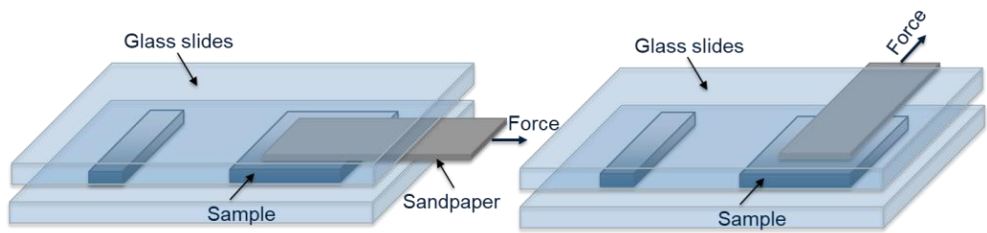

*Figure 8: Illustration of experiment setup for the surface scratch test.*

**4.3 Results and discussion**

The normal-incident reflectance spectra were collected on the same GaAs DBR at the same location with 0, 10, 20, 40, 60, 100, and 150 times of sandpaper scratching both laterally and vertically, which are shown in Figure 9 (a). First of all, no spectrum shift and no additional peaks were observed on the scratched sample. The stopband reflectance had a significant reduction of around 20% and 30% from the unscratched DBR to the DBR with 10 and 20 times of scratching respectively. This is expected as light scattering happens when the surface is no longer smooth and flat. With the further increase in the number of scratching up to 150, the reflectance of the stopband left side continues to decrease, while only a minor decrease was seen on the right side of the stopband. More importantly, a dip starts to appear in the middle of the stopbands with 40 or more times of scratching. It strongly suggests that the double-peak-shaped stopband of the GaAsP/Si DBR is due to the cross-hatch pattern on the DBR surface, which was reproduced with the GaAs DBR with 150 times of scratching as seen in Fig. 9 (b).

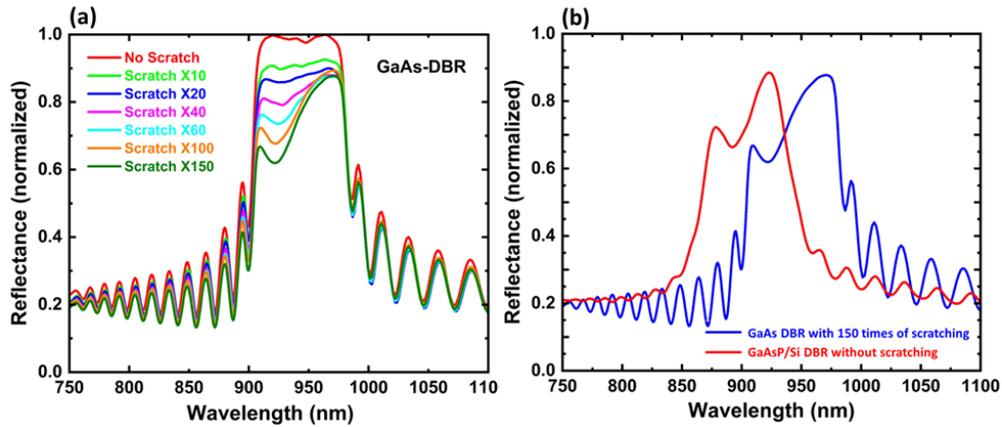

*Figure 9: (a) normal-incidence reflectance spectrum of the GaAs DBR with different times of scratching in both the lateral and vertical directions, and (b) comparison of the spectrum of the scratched GaAs DBR with 150 times of scratching and the GaAsP/Si DBR.*

AFM analysis was conducted on the GaAs DBR with 150 times of scratching to investigate and compare the surface roughness with graded GaAsP/Si DBR (Fig.

10). The AFM RMS roughness for the scratched GaAs DBR is 9.17 nm and 15.9 nm at the sample center and edge respectively. These values are similar to the RMS of graded GaAsP/Si DBR of 8.52 nm. High density of particles is seen on the surfaces along the scratches, which are the materials removed from the surface. The peak-to-valley height difference obtained from AFM results of graded GaAsP/Si DBR is 60 nm, which is much smaller than the stopband wavelength range of 850 – 960 nm. Therefore, destructive interference from this small surface roughness in the out-of-plane direction can be ignored. From Fig. 10, we can see that the lateral distance between the scratches is around one micron or less, similar to the average distance between the ridge and valley in the GaAsP/Si DBR surface cross-hatch of one to few microns, and the stopband wavelength range. Therefore, we believe is the in-plane cross-hatch feature size, i.e., the ridge/valley lateral distance, are more relevant to the destructive interference in this wavelength range.

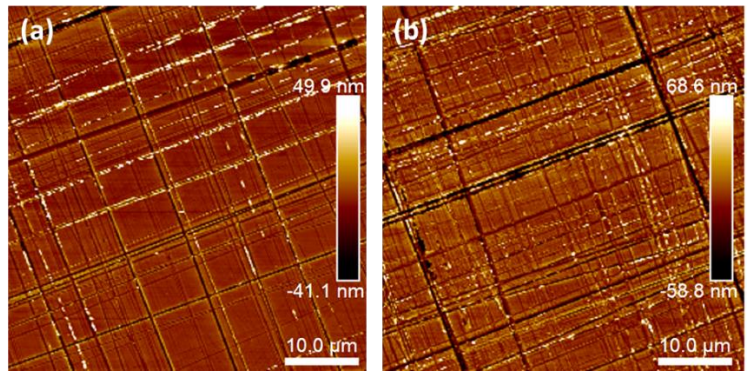

*Figure 10: AFM image of the GaAs DBR with 150 times of scratching at (a) the center of sample and (b) the edge of sample.*

### 4.4 Surface polishing test

Chemical mechanical polishing (CMP) was attempted on the GaAsP/Si DBR to reduce the cross-hatch pattern for a smoother surface. X-SEM results on the CMPed GaAsP/Si DBR showed that 500 nm was removed by CMP, including three full periods and a partial period of DBR. 1D TMM simulations show that 36 full DBR periods should still give an acceptable flat-top stopband. Both normal-incident reflectance spectrum and AFM images were collected on GaAsP/Si DBR with CMP, as shown in Fig. 11. A cross-hatch pattern was still observed as well after the CMP, but the RMS value of the CMPed GaAsP/Si DBR is 3.69 nm, compared to the 8.52 nm RMS roughness value of the GaAsP/Si without CMP. The reflectance spectrum shows that a flat stopband is achieved on the GaAsP/Si DBR after CMP with a maximum reflectance of 79.93 % and a stopband width of 75.44 nm. Fig. 11 (b) shows that CMP induced shallower and thinner surface scratches in random directions, compared to the deeper and wider scratches from the surface scratch test. Much fewer particles are observable on the CMPed surface.

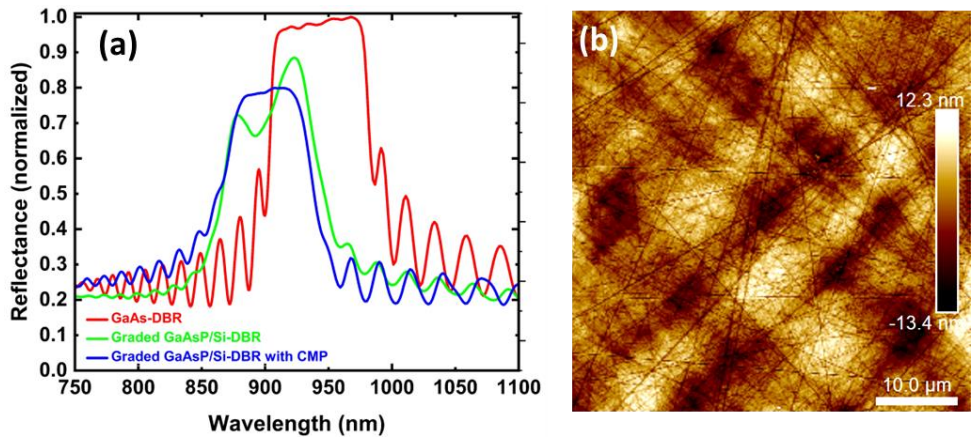

*Figure 11: Reflectance spectrum (a) of the CMPed GaAsP/Si DBR compared with the GaAs-DBR and the GaAsP/Si DBR without CMP; (b) AFM image of the GaAsP/Si DBR with CMP*

Overall, both the surface scratch test and the CMP test show that a defect-free and cross-hatch-free surface morphology is essential to the success of a DBR. In the previously studied GaAs/ART-Ge/Si surfaces, no cross-hatch was observed, likely due to the three-micron thick Ge growth on Si and the CMP step after the Ge growth before the GaAs top layer. More epitaxy and CMP development to reduce the surface cross-hatch of the GaAsP/Si substrates should be helpful.

## 5. CONCLUSION

Monolithically integrated 940 nm AlGaAs DBRs on graded GaAsP/Si substrates were studied and compared with the DBRs on the GaInAs/bulk-Ge substrates and bulk GaAs substrates. The surface of the GaAsP/Si DBRs have a low density of surface bumps as well as cross-hatch patterns. Normal-incidence reflectance spectra of the GaAsP/Si DBRs show double-peak-shaped stopbands. The surface scratch test and CMP test were conducted, which suggest that the cross-hatch pattern on the DBR surface is the root cause of the double-peak-shaped stopband. More R&D work in the buffer layer design, epitaxy growth and surface polishing are still needed to make graded GaAsP/Si substrates relevant to the VCSEL industry.

## ACKNOWLEDGE

Huawei Technologies, Canada is acknowledged for funding this project. Umicore N. V., Belgium, is acknowledged for providing the bulk Ge wafers with GaAs/GaPAs epitaxy layers and SiN back coating.